\documentclass[prl,showpacs,twocolumn]{revtex4}
\usepackage{graphicx} 

\begin{document}

\title{A Theoretical Search for the Optimum Giant Magnetoresistance}
\author{Tat-Sang Choy}
\author{Jian Chen}
\altaffiliation{Seagate Technologies, Minneapolis, MN 55435}
\author{Selman Hershfield}
\affiliation{Department of Physics and National High Magnetic Field 
Laboratory, University of Florida, Gainesville, FL 32611}

\date{\today}

\begin{abstract}
The maximum current-perpendicular-to-plane 
giant magnetoresistance is searched for
in magnetic multilayers made of Co, Ni, and Cu with disorder
levels similar to those found in room temperature experiments.
The calculation is made possible by a highly optimized 
linear response code, which uses the impurity averaged Green's
function technique and a 9-band per spin tight binding model.
Using simulated annealing, hundreds of different configurations
of the atomic layers are examined to find a maximum GMR of 450\% in 
ultrathin Ni/Cu superlattices.
\end{abstract}

\pacs{75.70.Pa, 72.25.Mk, 72.15.Gd}
\keywords{giant magnetoresistance, magnetic multilayers, optimization}

\maketitle

The phenomena of the giant magnetoresistance
(GMR) has been extensively investigated both
theoretically and experimentally since its 
discovery.\cite{discovery1,discovery2}
This interest stems from both fundamental questions about magnetism
and transport on the nanoscale
and also from technological applications like magnetic 
sensors.\cite{applications}
From this body of work
a basic understanding of the physics behind the GMR has emerged.
When a magnetic field causes the magnetic domains in a magnetic multilayer
or other GMR structure to change their
relative orientation, the resistance changes because 
the bulk and interface scattering rates depend on the electron
spin orientation.
While this is simple to explain, the actual details are quite complex.
Experimentally, it is known that the GMR is very sensitive to atomic composition
of the magnetic multilayers.
In cases adding a single monolayer 
can change the magnetoresistance.\cite{expsensitive1}
Film growth parameters
can similarly change the GMR by an order of magnitude.\cite{expsensitive2}
The sensitivity to atomic composition and film morphology as well as
GMR samples consisting of many atomic layers has made the theory of the
GMR challenging (for reviews see Refs. \onlinecite{theoryrev1} and
\onlinecite{theoryrev2}).

Nonetheless, much progress has been made in 
elucidating the physics of the GMR in general
and in explaining specific experiments.
Thus, in this paper we take a different approach.
Based on our work and those of other groups, we assume that we have
an accurate model of the GMR.  
\cite{dossmearing}
We then ask the question: 
what is the largest GMR one can obtain for a class of magnetic multilayers
with experimentally reasonable disorder parameters?
In other words, can one theoretically search the parameter space of 
a specific class of GMR structures and find the optimum or near optimum
atomic configuration?  This is a much more challenging
problem than computing the GMR for a specific GMR structure because one
must perform many individual GMR calculations in the search for the
optimum magnetoresistance.  
Indeed there have been very few calculations like this
in condensed matter physics. 
Franceschetti and Zunger demonstrated the use of simulated annealing
to optimize the band gap in semiconductor alloys and
superlattices. By searching as much as $10^4$ samples,
they found the maximum band gap configuration to 
be very complex.\cite{naturework}
Using a similar approach, \'{I}\~{n}iguez and Bellaiche optimized the
ground state structural properties and electromachanical response 
of perovskite alloys.\cite{perovskite}
This article presents one of the first 
attempts to optimize a transport property.

Optimizing the transport properties of nanostructures requires a delicate
balance between accuracy and speed. 
To ensure that the optimization results are valid, 
the model has to include the band structure
of the materials, the geometry of the nanostructure, and scattering effects 
caused by impurities, disorder, and temperature. 
On the other hand, the calculation of the transport property has to be
fast enough that the optimization 
can be finished in a reasonable amount time. 
To solve these difficulties, we have developed a 
highly optimized parallel tight-binding code to compute electron 
transport in nanostructures using the impurity averaged Green function
technique and linear response theory.
While the tight binding model we use is not as accurate as 
other techniques such as density functional theory,\cite{dftwork}
it does allow us to include realistic
band structure efficiently.  Similarly, the impurity averaged Green's
function technique allows us to include different kinds of interface and
bulk scattering without worrying about the placement of individual
defects.  

The tight binding Hamiltonian we consider is
\begin{equation}
   H = \sum _{{\mathbf r},j;{\mathbf r'},j'}
       H_{{\mathbf r},j;{\mathbf r'},j'}
       c^\dag_{{\mathbf r},j}
       c_{{\mathbf r}',j} ,
\label{hamiltonian}
\end{equation}
where ${\mathbf r}$ and ${\mathbf r}'$ are sites on the lattice.  
The indices $j$ and $j'$ denote
one of the one {\it s}, three $p$, or five {\it d} levels 
and also the spin orientation.
Within a tight binding model, the disorder comes from variations in the
matrix elements.  
In the impurity averaged green function technique, these variations 
are treated statistically by averaging over an ensemble of
different microscopic configurations of the disorder with the same 
macroscopic properties, e.g., density of impurities.
Denoting this average by angular brackets, our unperturbed Hamiltonian
is $H_o = \langle H \rangle$.  The deviation from this average for a 
particular configuration of the disorder is given by
$\delta H = H - H_o$.

Scattering is included through the choice of the self-energy.
We use the self-consistent Born approximation with uncorrelated
disorder fluctuations.
For energy $\omega $
the retarded and advanced self-energies, $\Sigma ^{\textrm R/A}$, are
expressed in terms of the retarded and advanced green functions,
$G^ {\textrm R/A}$ as
\begin{equation}
   \Sigma ^{\textrm R/A}_{{\mathbf r},j;{\mathbf r},j} (\omega ) =
   \langle (\delta H_{{\mathbf r},j;{\mathbf r},j})^2 \rangle 
   G ^{\textrm R/A}_{{\mathbf r},j;{\mathbf r},j} (\omega ),
   \label{selfenergy}
\end{equation}
where the green function is in turn expressed in terms of the self-energy via
\begin{equation}
   [\omega - H_o - \Sigma ^{\textrm R/A}(\omega )]
   G ^{\textrm R/A}(\omega ) = {\mathbf 1} .
   \label{dysoneq}
\end{equation}
This last equation is a matrix equation, where the ${\mathbf r}$ and $j$ indices have
been suppressed.

Within the linear response theory, the conductivity is computed via the 
Kubo formula.  For the geometry with the current flowing perpendicular to
the planes (CPP) it is necessary to include the vertex
correction, $\Gamma$, which 
is given by the self-consistent equation, 
\begin{equation}
   {\mathbf \Gamma} _{{\mathbf r},j;{\mathbf r},j}  =
   \langle (\delta H_{{\mathbf r},j;{\mathbf r},j})^2 \rangle 
   \left( G^{\textrm R} ({\mathbf J} + {\mathbf \Gamma} ) G^{\textrm A} 
   \right)\Big|_{{\mathbf r},j;{\mathbf r},j},
   \label{vertexcorrection}
\end{equation}
where ${\mathbf J}$ is the current operator,
\begin{equation}
   {\mathbf J}_{{\mathbf r},j;{\mathbf r'},j'}  =
       -i ({\mathbf r} - {\mathbf r'})
       H_{{\mathbf r},j;{\mathbf r'},j'} .
   \label{totalcurrent}
\end{equation}
A conserving approximation for the conductivity at ${\mathbf r}$ is 
\begin{eqnarray}
   \sigma _{\alpha\beta}(\mathbf r) &=&
   \frac {e^2}{h}
   \mbox{Tr}\{
J_\alpha ({\mathbf r}) G^{\textrm R} (J_\beta +\Gamma _\beta ) G^{\textrm A } \\
   &-& \frac 12 J_\alpha ({\mathbf r}) G^{\textrm R } J_\beta G^{\textrm R }
   - \frac 12 J_\alpha ({\mathbf r}) G^{\textrm A } J_\beta G^{\textrm A } \} ,
   \nonumber
   \label{conductivity}
\end{eqnarray}
where ${\mathbf J}({\mathbf r})$ is the local current density operator
at ${\mathbf r}$.
The sum of ${\mathbf J}({\mathbf r})$ over position times the lattice unit cell
volume is equal to the current operator in 
Eq. (5).
In the CPP geometry this conductivity will be constant
because the current density is independent of position.

Letting $\sigma _{\textrm F}$ be the conductivity 
when the magnetic layers are aligned
ferromagnetically and $\sigma _{\textrm AF}$ be the conductivity when
adjacent magnetic layers are aligned antiferromagnetically,
the giant magnetoresistance is defined as
${\textrm GMR} = (\sigma _{\textrm F} 
- \sigma _{\textrm AF})/\sigma _{\textrm AF}$.
Note that in practice the magnetic layers will not necessarily
be aligned antiferromagnetically at zero field; however, our simulation
does not compute the magnetic domain structure, but only the conductivity.
In particular some of the ultrathin magnetic layers which we find
with large GMR could very well be ferromagnetically coupled.\cite{coupling}
It is possible to achieve antiparallel alignment even with ferromagnetic
coupling by going to a spin valve geometry with only two magnetic
layers where one of the layers is exchange biased.\cite{spinvalve}

An optimization study involves many GMR calculations, 
therefore, the speed for each GMR calculation is crucial.
The trace in Eq. (6) is over all positions
and atomic levels; however, since the system is periodic, 
it can be broken up into 
a sum in k-space and a sum within a unit cell. 
Thus, the matrices are of order $9N$ for each spin, 
where $N$ is the number of atoms in a unit cell.
At first glance, the matrix products and inverses in Eqs. (3), (4) and (6) 
are all $O(N^3)$ operations, where $N$ is the number of atoms in the unit cell. 
In addition, both Eqs. (3) and (4) are self-consistent equations. 
Performing optimization with a self-consistent $O(N^3)$ implementation
is estimated to take about 5 CPU-years, which is impractical.
To overcome this, we have developed an $O(N)$ algorithm to solve 
Eqs. (3) and (6), and an $O(N^2)$ algorithm to solve 
Eq. (4) directly without the need for iteration.
For $N=50$, this algorithm is more than 100 times faster than 
the $O(N^3)$ algorithm.
Running on a Linux cluster of five dual 866MHz Pentium III CPU workstations,
it takes about 20 minutes
for each GMR calculation and about two weeks for an optimization.

The optimization technique we use is simulated annealing.
Since not every configuration qualifies as a GMR structure,
the optimization is performed in the constrained subspace
consisting of a unit cell with two magnetic layers separated
by two non-magnetic spacer layers.
After an arbitrarily chosen initial unit cell, 
each subsequent configuration is generated by one of the
following Monte Carlo moves:
(i) inserting a monolayer,
(ii) removing a monolayer, or
(iii) changing the composition of a monolayer.
If the GMR of the new configuration is higher than the previous
configuration, then it is accepted.  On the other hand,
if the new GMR is lower, it is only accepted with a probability
of $\exp (\Delta\mbox{GMR}/T)$, where $\Delta\mbox{GMR}$ is the
change in the GMR and $T$ is the simulated annealing temperature.
The process is continued as the annealing temperature is gradually
lowered until there is no further change in the configuration
and a final near-global maximum in the GMR is attained.

In the following, we show results of a study that looks for the optimal
configuration of superlattices made with Co, Ni, and Cu in conditions
similar to GMR experiments at room temperature.  To make the study more
manageable, we only consider fcc (111) lattices.  The minimum thickness of
each layer is limited to two monolayers in order to avoid complications
such as pin holes.  The tight binding parameters in Eq. (1) are
obtained from fits to density-functional calculations where up to 
second-nearest-neighbor hoping energies are included.

Two types of scattering are included: spin-independent scattering 
and spin-dependent scattering.  The spin-independent scattering
represents the structural disorder and is included in the calculation
by uniformly shifting all the orbital levels at a site:
$\langle (\delta H_{{\mathbf r},j;{\mathbf r},j})^2 \rangle 
= 0.2 {\textrm eV}^2$,
which corresponds to a Cu resistivity of 3.1 $\mu \Omega {\textrm cm}$, 
about twice the value of clean Cu at room temperature.
The spin-dependent scattering is assumed to be due to inter-diffusion
of atoms near the interface.  It is modeled by the shifting of the
{\it d}-levels only, since the {\it d}-levels are the primary difference
in the tight binding parameters for these elements. 
The shift in the {\it d}-levels are determined by fits to 4-atom
supercell density-functional calculations of an impurity in three host atoms.
We then assume an interface mixing percentage of 10\% in each of the
two monolayers on either side of the interface.
The value of $\langle (\delta H_{{\mathbf r},j;{\mathbf r},j})^2 \rangle$ 
for the majority and minority {\it d}-states is equal to 0.29 and 
0.84 ${\textrm eV}^2$ at the Co/Cu interface, and 0.03 and 
0.11 ${\textrm eV}^2$ at the Ni/Cu interface.
Using these parameters, 
the GMR for a [Co$_3$/Cu$_3$/Co$_3$/Cu$_3$] superlattice
is 36\%, 
which is in the range of observed CPP-GMR  at room temperature.\cite{cocuexp}
Our spin-dependent scattering does not include spin flip scattering,
which can be important for short spin flip scattering lengths.\cite{valetfert}

\begin{figure}
\includegraphics[width=8.6cm]{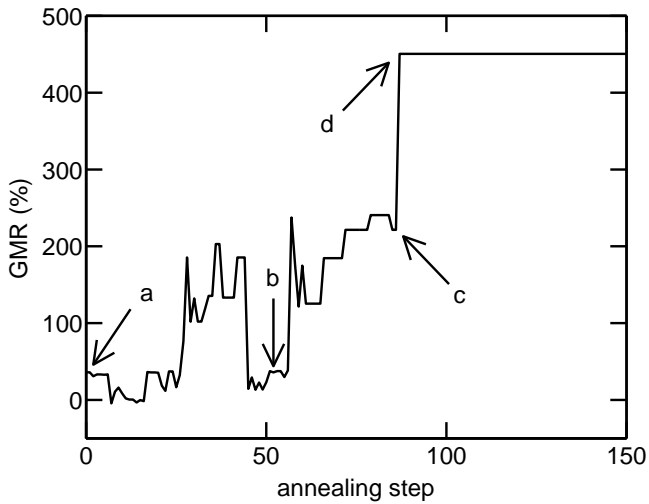}
\caption{GMR as a function of the simulated annealing step.
The four marked configurations and their GMR values are:
(a) [Co$_3$/Cu$_3$/Co$_3$/Cu$_3$] $36\%$, (b) [Ni$_2$/Cu$_3$/Ni$_2$/Cu$_2$]
$221\%$, (c) [Ni$_2$/Cu$_2$/Ni$_2$/Co$_1$/Cu$_2$] $38\%$, and (d)
[Ni$_2$/Cu$_2$/Ni$_2$/Cu$_2$] $450\%$. Notice that the GMR is
very sensitive to the configuration. Both (b) and (c) differ from (d)
by only one monolayer. 
}
\end{figure}

The giant magnetoresistance 
for a simulated annealing run is shown in Fig. 1.
The initial configuration has been chosen to be the superlattice
[Co$_3$/Cu$_3$/Co$_3$/Cu$_3$],
with a unit cell
made of four layers, 
each consisting of three monolayers of either Co or Cu.
The particular choice of initial configuration
does not effect the final result for this calculation.
As the annealing temperature $T$ decreases from its initial
value of $60\%$, the GMR becomes larger. 
The GMR does not increase monotonically because 
a superlattice that reduces the GMR is accepted with 
a probability set by the Boltzmann factor, which is
crucial for escaping from local maxima in the GMR. 
As the annealing temperature is lowered, the chance of leaving
a local maximum becomes smaller and smaller.  At step 89 in this run,
the GMR jumps to $450\%$ and 
no other configurations are accepted.
The atomic configuration for the highest
GMR (d) is [Ni$_2$/Cu$_2$/Ni$_2$/Cu$_2$].
This is a very surprising result because most
room temperature studies to date have GMR values less 
than $100\%$.\cite{cocuexp,theoryrev2,msurev}
From Fig. 1 one can see that
the GMR is very sensitive to the atomic configuration.
For example, superlattices (b) and (c) have GMR values
of 221\% and 38\%, respectively, yet they differ by only one
monolayer from the optimal GMR of superlattice (d), 450\%.

\begin{figure}
\includegraphics[width=8.6cm]{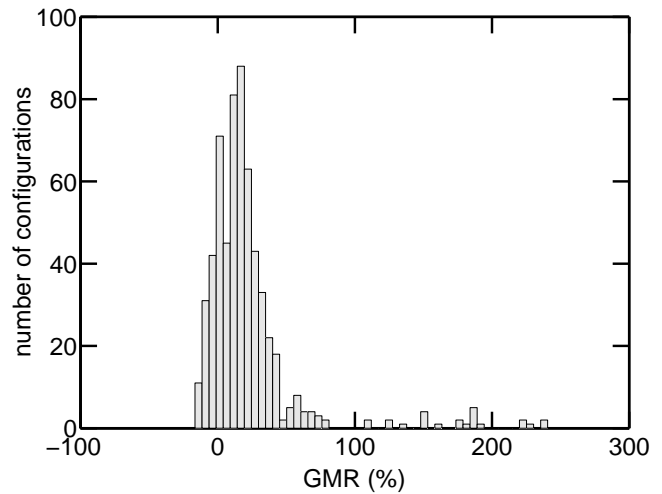}
\caption{Histogram of the GMR in the configuration space
for 600 randomly generated configurations.
The average GMR is $22\%$, and the standard deviation is $36\%$.
About $97\%$ of the configurations are located 
near the main peak with GMR values less than 100\%, 
while 3\% of the configurations 
are many standard deviations away from the main peak
with GMR values greater than 100\%.
These large GMR configurations are mostly ultrathin Ni/Cu superlattices.
}
\end{figure}

To get a better picture of the GMR in configuration
space, we have calculated the GMR for 600 randomly generated configurations.
The result is shown in Fig. 2 as a histogram. 
The average GMR 
is $22\%$, and the standard deviation is $36\%$.
There are two groups of configurations. 
About $97\%$ of the configurations are located near the main peak.
They have a GMR less than $100\%$. On the other hand, about $3\%$ of
the configurations have GMR many standard deviations away from the main peak. 
They scatter in the range from $100\%$ to $450\%$.
Most of the large GMR configurations are ultrathin Ni/Cu superlattices.
Only two of the large GMR configurations contain Co.
Since the large GMR configurations are rare, they may have been overlooked.

What causes the GMR to be large in some multilayers but not others?
The dominant source of scattering in these transition metals 
is due to {\it s}-{\it d} hybridization.  
The {\it d}-levels in the majority spin of Ni, Co, and Cu are below the
Fermi level; however, there is still a finite {\it d}-density of states at the
Fermi level due to both {\it s}-{\it d} hybridization and disorder.
In particular disorder can substantially smear the {\it d}-density of states,
increasing the value at the Fermi level.\cite{dossmearing}  
Since the {\it d}-levels in Ni and Cu are
more well matched than those in Co and Cu, the 
disorder due to interface mixing is smaller in Ni/Cu multilayer than 
in a Co/Cu multilayer. 
Thus, the GMR is greater in a Ni/Cu multilayer 
for the same level of interface mixing because
scattering via the {\it s}-{\it d} hybridization is less.

\begin{figure}
\includegraphics[width=8.6cm]{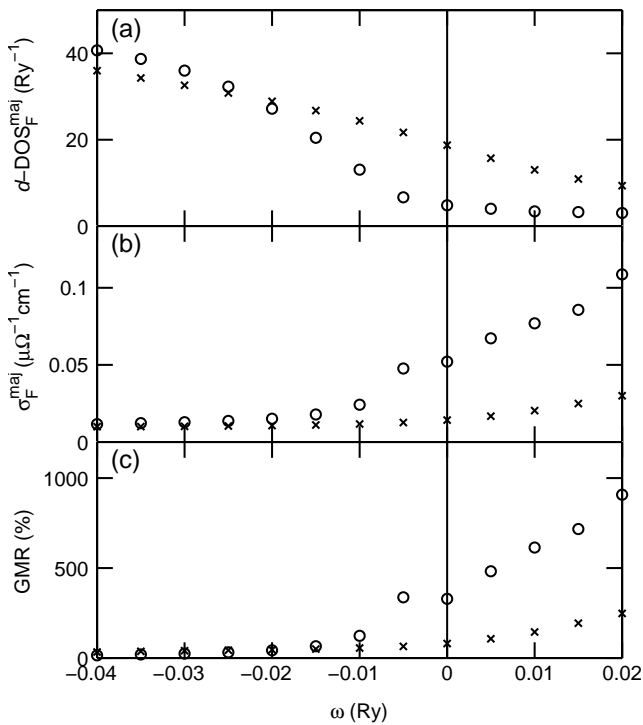}
\caption{
Comparison of [Ni$_2$/Cu$_2$/Ni$_2$/Cu$_2$] (circles) and
[Co$_2$/Cu$_2$/Co$_2$/Cu$_2$] (crosses) superlattices.
(a) At the Fermi level the {\it d}-density of states
for the majority spin when the magnetic layer aligns ferromagnetically,
{\it d}-DOS$_{\textrm F}^{\textrm maj}$,
is higher in the Co/Cu sample, leading to higher scattering.
(b) The conductivity
for the same spin channel,
$\sigma_{\textrm F}^{\textrm maj}$, in Co/Cu is thus lower than that in Ni/Cu. 
Since near the Fermi energy the conductivities for the minority spin
and the antiferromagnetically aligned magnetic layers are
relatively small and flat compared with $\sigma_{\textrm F}^{\textrm maj}$, 
the GMR (c) resembles $\sigma_{\textrm F}^{\textrm maj}$.
}
\end{figure}

To see this more clearly, in Fig. 3 we take a representative
Ni/Cu sample and a representative Co/Cu sample.
The two samples have the same bulk disorder, but the 
Ni based multilayer actually has more interface mixing (10\%)
than the one containing Co (5\%) to allow the curves to
fit clearly on the same plots.
As seen in Fig. 3(a),
the {\it d}-density of states is smaller in the Ni sample.
The {\it d}-density of states correlates well with the
larger conductivity shown in Fig. 3(b)
and the larger GMR shown in Fig. 3(c).
The fact that there is a simple explanation for this large
GMR indicates that it is a robust effect and not strongly dependent
on the many assumptions which we have made.
Also, it says that for comparable coupling between layers and
disorder, the Ni/Cu multilayers will have a larger GMR than
the Co/Cu ones.

In this paper we have performed a simulated annealing search for the
optimum or maximum giant magnetoresistance by varying the atomic
composition of multilayers composed of Ni, Co, and Cu.
The disorder parameters in our calculation
were chosen to simulate room temperature experimental conditions.
After hundreds of GMR calculations for different atomic
configurations, a surprisingly large GMR was found for
ultrathin
Ni/Cu multilayers.  The origin of this large GMR of 450\% was
found to be the fact that the {\it d}-levels in Ni and Cu are relatively
close in energy.  
Thus, although we have made a number of approximations in order
to perform our calculation in a reasonable amount of time,
the large GMR appears to be a robust effect.
We believe that this
work is part of a large class of problems in tailoring
transport properties of nanostructures to optimize an observable
such as the magnetoresistance.

\begin{acknowledgments}
We would like to thank Fred Sharifi for useful discussions.
This work is supported by DOD/AFOSR F49620-1-0026,
NSF DMR9357474, and the Research Corp.
\end{acknowledgments}


\end{document}